\documentclass[
reprint,
superscriptaddress,
 amsmath,amssymb,
]{revtex4-2}

\usepackage{graphicx}% Include figure files
\usepackage{dcolumn}% Align table columns on decimal point
\usepackage{bm}% bold math
\begin{document}

\title{Antiferromagnetic order in MnBi$_2$Te$_4$ films grown on Si(111) by molecular beam epitaxy}
\author{N. Liu}
%\email{nan.liu@physik.uni-wuerzburg.de}
\affiliation{Faculty for Physics and Astronomy (EP3), Universit$\ddot{a}$t W$\ddot{u}$rzburg, Am Hubland, D-97074, W$\ddot{u}$rzburg, Germany}
\affiliation{Institute for Topological Insulators, Am Hubland, D-97074, W$\ddot{u}$rzburg, Germany}
\author{S. Schreyeck}
\affiliation{Faculty for Physics and Astronomy (EP3), Universit$\ddot{a}$t W$\ddot{u}$rzburg, Am Hubland, D-97074, W$\ddot{u}$rzburg, Germany}
\affiliation{Institute for Topological Insulators, Am Hubland, D-97074, W$\ddot{u}$rzburg, Germany}
\author{K. M. Fijalkowski}
\affiliation{Faculty for Physics and Astronomy (EP3), Universit$\ddot{a}$t W$\ddot{u}$rzburg, Am Hubland, D-97074, W$\ddot{u}$rzburg, Germany}
\affiliation{Institute for Topological Insulators, Am Hubland, D-97074, W$\ddot{u}$rzburg, Germany}
\author{M. Kamp}
\affiliation{Physikalisches Institut and R$\ddot{o}$ntgen-Center for Complex Material Systems (RCCM), Fakult$\ddot{a}$t f$\ddot{u}$r Physik und Astronomie, Universit$\ddot{a}$t W$\ddot{u}$rzburg, Am Hubland, D-97074 W$\ddot{u}$rzburg, Germany}
\author{K. Brunner}
\affiliation{Faculty for Physics and Astronomy (EP3), Universit$\ddot{a}$t W$\ddot{u}$rzburg, Am Hubland, D-97074, W$\ddot{u}$rzburg, Germany}
\affiliation{Institute for Topological Insulators, Am Hubland, D-97074, W$\ddot{u}$rzburg, Germany}
\author{C. Gould}
\affiliation{Faculty for Physics and Astronomy (EP3), Universit$\ddot{a}$t W$\ddot{u}$rzburg, Am Hubland, D-97074, W$\ddot{u}$rzburg, Germany}
\affiliation{Institute for Topological Insulators, Am Hubland, D-97074, W$\ddot{u}$rzburg, Germany}
\author{L. W. Molenkamp}
%\email{Laurens.Molenkamp@physik.uni-wuerzburg.de}
\affiliation{Faculty for Physics and Astronomy (EP3), Universit$\ddot{a}$t W$\ddot{u}$rzburg, Am Hubland, D-97074, W$\ddot{u}$rzburg, Germany}
\affiliation{Institute for Topological Insulators, Am Hubland, D-97074, W$\ddot{u}$rzburg, Germany}

\date{\today}

\begin{abstract}
MnBi$_2$Te$_4$ has recently been predicted and shown to be a magnetic topological insulator with intrinsic antiferromagnetic order. However, it remains a challenge to grow stoichiometric MnBi$_2$Te$_4$ films by molecular beam epitaxy (MBE) and to observe pure antiferromagnetic order by magnetometry. We report on a detailed study of MnBi$_2$Te$_4$ films grown on Si(111) by MBE with elemental sources. Films of about 100\,nm thickness are analyzed in stoichiometric, structural, magnetic and magnetotransport properties with high accuracy. High-quality MnBi$_2$Te$_4$ films with nearly perfect septuple-layer structure are realized and structural defects typical for epitaxial van-der-Waals layers are analyzed. The films reveal antiferromagnetic order with a Néel temperature of 19\,K, a spin-flop transition at a magnetic field of 2.5\,T and a resistivity of 1.6\,m$\Omega$$\cdot$cm. These values are comparable to that of bulk MnBi$_2$Te$_4$ crystals. Our results provide an important basis for realizing and identifying single-phase MnBi$_2$Te$_4$ films with antiferromagnetic order grown by MBE.
\end{abstract}

\maketitle

MnBi$_2$Te$_4$ is a magnetic topological insulator with antiferromagnetic order \cite{2019SciAdv,2019Nature-MBT}. It is a stoichiometric compound with septuple layers (7Ls) consisting of Te-Bi-Te-Mn-Te-Bi-Te, which are van-der-Waals (vdW) bound along the c axis. The hexagonal Mn layers of neighboring 7Ls have antiparallel out-of-plane spins resulting in A-type antiferromagnetic order. Bulk MnBi$_2$Te$_4$ has a Néel temperature of 25\,K and a spin-flop transition at a magnetic field of about 3.5\,T \cite{2019Nature-MBT,Hosono-Sci.adv,Bulk-MBT}. Bulk crystal properties of MnBi$_2$Te$_4$ have been proposed to belong to a new classification of antiferromagnetic topological insulators with so called S symmetry \cite{2019SciAdv,2010AFMTI}. These novel symmetry properties may result in new realizations of topological properties like the quantized magnetoelectric effect and axion physics. The most studied magnetic topological insulator in the past few years are (Bi,Sb)$_2$Te$_3$ films, doped with Cr or V, which have been used to achieve the quantum anomalous Hall effect (QAHE) \cite{2010QAHE,2013QAHE,2014QAHE,2015QAHE,2017QAHE}. This material has a Curie temperature of about 20\,K, but a properly insulating bulk has so far been realized only at mK temperatures \cite{2018QAHE1,2018QAHE2,corbino-paper}. The QAHE has also been proposed for thin films of MnBi$_2$Te$_4$ consisting of an odd number of 7Ls \cite{2019SciAdv}. Recent magnetotransport studies of Hall bar like structures fabricated out of MnBi$_2$Te$_4$ flakes exfoliated from bulk crystals support these expectations, although a perfect QAHE and dissipationless edge channel transport at zero magnetic field is not yet achieved \cite{MBT-QAHEScience,2020YYWang}. 

Molecular beam epitaxy (MBE) of stoichiometric MnBi$_2$Te$_4$ promises well-controlled thin films on wafer-sized substrates, as well as surface protection by a capping layer. The development of MBE methods for MnBi$_2$Te$_4$, however, has started only recently \cite{MBE-KeHe, MBE-EP7,PRM-MBE,MBT-AHE1,MBT-AHE2,MBT-transport,MBT-AHE}. MBE growth is a challenge due to the competition between vdW 7Ls growth and many other, closely related phases in the (Bi,Mn)Te material system \cite{MBE-KeHe, MBE-EP7,PRM-MBE,MBT-AHE1,MBT-AHE2}. In magnetometric SQUID measurements of thin epitaxial films, ferromagnetic and (super)paramagnetic contributions of other phases obscure the antiferromagnetic response \cite{MBT-transport,MBT-AHE}. The Hall resistivity of films indicates two anomalous Hall components suggesting antiferromagnetic order but the influence of different phases and types of carriers is unclear \cite{MBT-AHE1,MBT-AHE2,MBT-transport,MBT-AHE}. Thus, the MBE growth of single-phase MnBi$_2$Te$_4$ films and a magnetometric proof of antiferromagnetic order characterized by a nearly vanishing magnetization at low magnetic field  is highly desirable.

In this work, we report on a detailed study of structural, magnetic, and magnetotransport properties in MnBi$_2$Te$_4$ films grown by MBE. A film thickness of about 100\,nm allows us to analyze the structural and magnetic properties with high accuracy. By adjusting the growth parameters, we obtain high-quality MnBi$_2$Te$_4$ films with nearly perfect 7Ls structure, as confirmed by x-ray diffraction (XRD), scanning transmission electron microscopy (STEM) and energy dispersive x-ray spectroscopy (EDS) measurements. In optimized MnBi$_2$Te$_4$ films, the antiferromagnetic ordering is identified by SQUID and transport properties, both of which are comparable to bulk single crystals.

\begin{figure*}
\centering\includegraphics[width=15 cm]{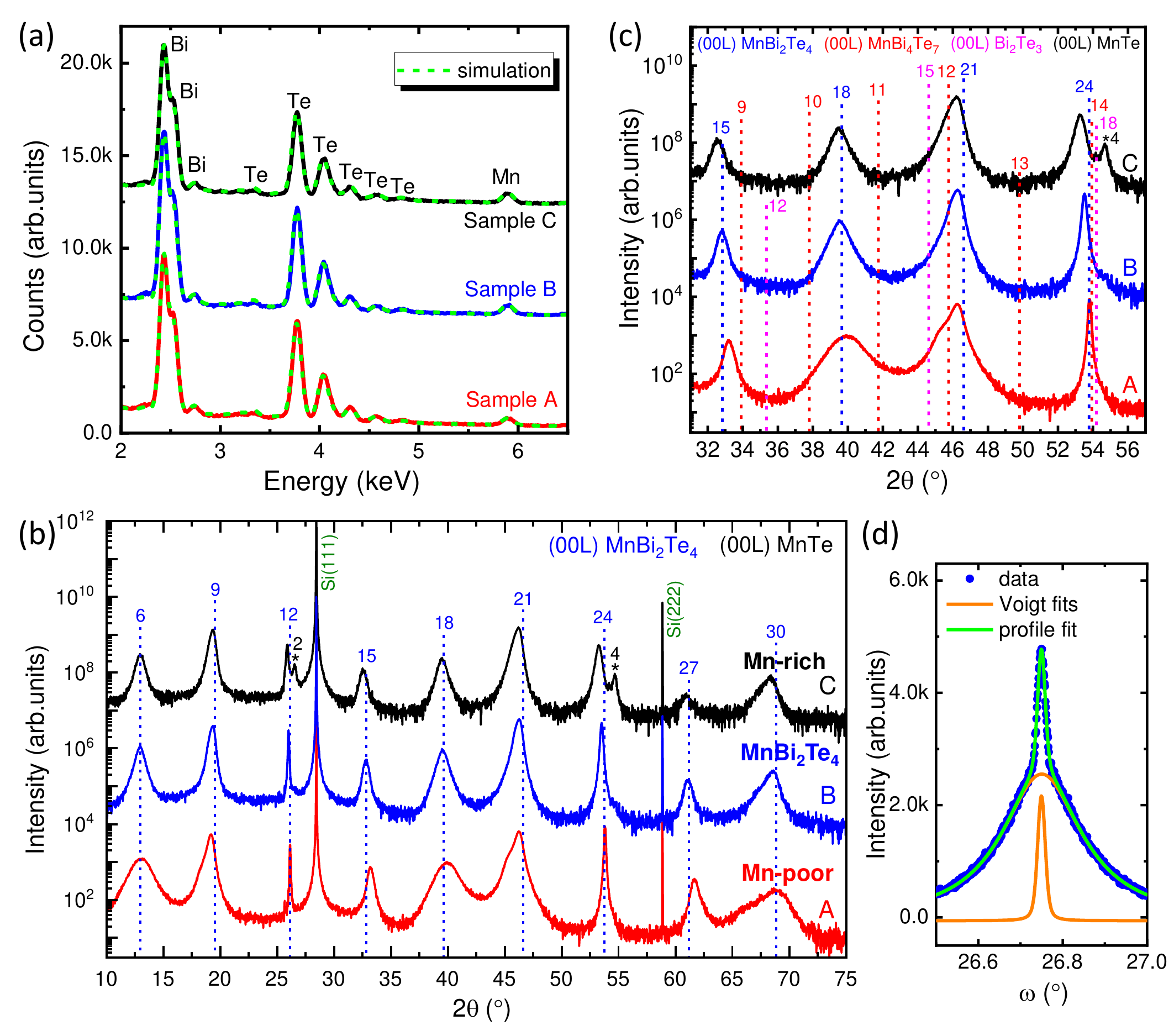}
\caption{(a) EDS measurements of samples A, B and C with the nominal Mn/Bi flux ratios being 0.31, 0.38 and 0.45, respectively. Simulation curves shown in green dashed lines are well fitted to the experiment data. Subsequent curves offset 6\,k counts for clarity. (b) Full range and (c) partial range $\theta$-2$\theta$ XRD scans of the same three samples A, B and C. The colored numbers L, dashed lines or asterisks mark the (00L) reflection positions of the following phases and (LLL) reflections of the Si substrate: MnBi$_2$Te$_4$ (blue), MnBi$_4$Te$_7$ (red), Bi$_2$Te$_3$ (pink), MnTe (black), Si (dark green). Subsequent curves offset by a factor of 10$^3$ for clarity. (d) Rocking curve scan of the (00\underline{24}) peak of sample B, the profile fit (green line) is the superposition of two Voigt fits (orange lines). 
}
\end{figure*}

Our MnBi$_2$Te$_4$ films are grown on hydrogen passivated 2-inch undoped Si(111) wafers in an MBE system with a base pressure of 1$\times$10$^{-10}$ mbar. Prior to MBE growth, the Si(111) substrates are dipped in 50\% HF solution and immediately transfered into a \textit{N}$_2$ glove box for clamp mounting on a PBN diffuser plate. These are then loaded into the MBE system without breaking the \textit{N}$_2$ protective atmosphere. During growth, the temperatures of the substrates are in the range from 220\,\textcelsius\, to 270\,\textcelsius, and high-purity elemental Bi (99.9999\%), Mn (99.9998\%), and Te (99.9999\%) sources are co-evaporated. The nominal beam fluxes of the sources are calculated based on the corresponding beam equivalent pressures (BEPs) measured by a Bayard-Alpert ion gauge close to the substrate position \cite{Flux-calibration}. The epitaxy process is carried out under Te-rich conditions with a BEP of about 5$\times$10$^{-8}$ Torr.

We now discuss three representative samples A (Mn-poor), B (Mn-ideal) and C (Mn-rich) to compare the basic characterization results. The nominal Mn/Bi flux ratios of samples A, B and C are 0.31, 0.38, 0.45, and the film thicknesses are 76 nm, 110 nm and 96 nm, respectively. As a first characterization experiment, we perform EDS measurements on the samples to determine their elemental compositions. As shown in Fig.\,1(a), Bi, Mn, and Te peaks are observed in the spectra of all three samples. For sample B, the atomic contents of Mn, Bi and Te elements are determined by simulation as 13.5 at.\%, 29.0 at.\% and 57.5 at.\%, respectively. This result is consistent with the stoichiometric compound MnBi$_2$Te$_4$ (i.e. Mn: 14.3 at.\%, Bi: 28.6 at.\% and Te: 57.1 at.\%), to within the measurement accuracy of a few at.\%. For the Mn-poor sample A we find Mn: 11 at.\%, Bi: 29 at.\% and Te: 60 at.\%, and finally, for the Mn-rich sample C we have Mn: 16 at.\%, Bi: 28 at.\% and Te: 56 at.\%. 

Fig.\,1(b) shows large scale $\theta$-2$\theta$ XRD scans of the same three samples. Sample B exhibits only the (00L) reflections of MnBi$_2$Te$_4$ at angles close to literature values \cite{Hosono-Sci.adv,MBT-peak}, plus the (111) and (222) reflections of the Si substrate. The full width half maximum (FWHM) 0.175$^\circ$ of the (00\underline{24}) peak is approximately twice the Scherrer limit for the given film thickness. The rocking curve ($\omega$ scan, see Fig.\,1(d)) of this peak reveals a relatively narrow FWHM\,=\,0.23$^\circ$, which is comparable to MBE grown vdW layers such as Bi$_2$Te$_3$ \cite{BTrockingcurve}. In addition, on top of this peak we observe a very sharp structure with a FWHM\,=\,0.022$^\circ$, which we take to indicate that some regions of 7Ls are highly oriented to the substrate. Overall, the results presented suggest that we have grown a single phase MnBi$_2$Te$_4$ layer with well-oriented 7Ls structure. 

A detailed comparison of samples A, B and C can be made based on the zoomed-in $\theta$-2$\theta$ diffractogram in Fig.\,1(c). The Mn-rich sample C exhibits an additional (004) peak of MnTe in NiAs-structure \cite{NiAs}, indicating the presence of MnTe inclusions. Some peaks assigned to MnBi$_2$Te$_4$ are shifted and broadened, indicating that the quality of the film is lower than that of sample B. For the Mn-poor sample A, the narrow peak at 2$\theta$ about 54$^\circ$ is shifted close to the position of the (00\underline{14}) peak in MnBi$_4$Te$_7$. The layer peaks of the (00\underline{15}), (00\underline{18}), and (00\underline{21}) reflections of MnBi$_2$Te$_4$ shift and broaden considerably as compared to sample B. Weak shoulders appear at the positions of the (00\underline{10}), (00\underline{11}), and (00\underline{12}) reflections of MnBi$_4$Te$_7$. These features indicate a mixture of MnBi$_2$Te$_4$ and MnBi$_4$Te$_7$ phases, i.e. intercalated Bi$_2$Te$_3$ quintuple layers (5Ls) in sample A. 

As a conclusion of the XRD results, the stoichiometric MnBi$_2$Te$_4$ sample B indicates a nearly perfect single-phase layer, while the Mn-rich (C) and Mn-poor (A) samples show MnBi$_2$Te$_4$ as the main phase, but respectively with MnTe and Bi$_2$Te$_3$ inclusions. The (00\underline{24}) peak is suitable to analyze the structural coherence and tilt in MnBi$_2$Te$_4$ films, since, given the overlapping reflections for all three phases, i.e. MnBi$_2$Te$_4$, MnBi$_4$Te$_7$, and Bi$_2$Te$_3$, at nearly the same Bragg angle, it is less affected by broadening. The MnBi$_2$Te$_4$ stoichiometry can be mainly verified by the positions, shapes and shoulders of peaks like (00\underline{21}) and the absence of peaks of other phases. 

\begin{figure}
\includegraphics[width=8.5 cm]{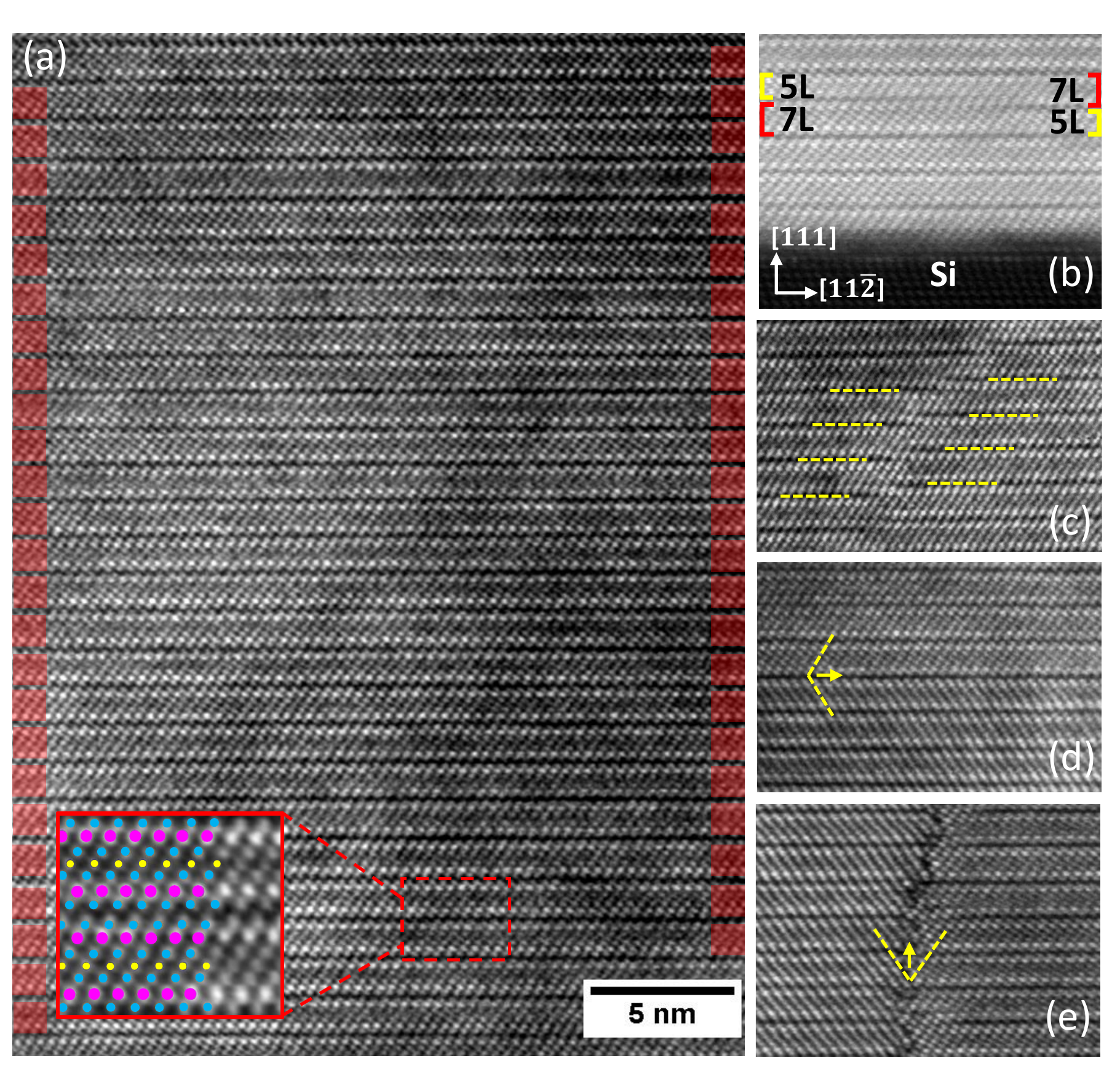}
\caption{Scanning transmission electron microscopy (STEM) cross-sectional images of sample D. (a) A representative cross-sectional image of a large area of pure 7Ls. Red blocks on both sides are drawn to highlight the 7Ls. The inset shows a close-up view of two adjacent 7Ls with higher resolution, in which the atoms are marked as: Te-blue, Bi-pink and Mn-yellow. (b) and (c) Splicing and merging of a 7L and a 5L and of 7Ls of twin domains with translation along c-axis, respectively. 7L and 5L are marked in red and yellow at the edges. The yellow dashed lines mark the positions of the vdW gaps between adjacent 7Ls. (d) and e) Rotational twins with boundaries parallel and perpendicular to layers marked by yellow arrows, respectively. 
}
\end{figure}

Figure 2 shows representative STEM cross-section images of sample D (nominally identical to sample B) measured along [11$\overline{2}$]. Perfect 7Ls stacking, without any defect, can be observed over a large area, as shown in Fig.\,2(a). In a close-up image with higher resolution (inset of Fig.\,2(a)), the atomic-scale structure of two adjacent 7Ls is clearly seen. In addition, even in this highly crystalline sample, we observe various types of defects in some regions of the epitaxial layer, as shown in Fig.\,2(b-e): we observe splicing and merging of a 7L and a 5L (b), of two 7Ls of twin domains with translation along c-axis (c), boundaries of 60$^\circ$ rotational twins parallel (d) and vertical (e) to the Si(111) surface. All these types of defects are typical for epitaxial vdW layer structures and have previously been observed in 5Ls structure of Bi$_2$Se$_3$ and Cr-doped (Bi,Sb)$_2$Te$_3$ \cite{EP3-AM, EP3-CEC}. 

\begin{figure}
\includegraphics[width=8.5 cm]{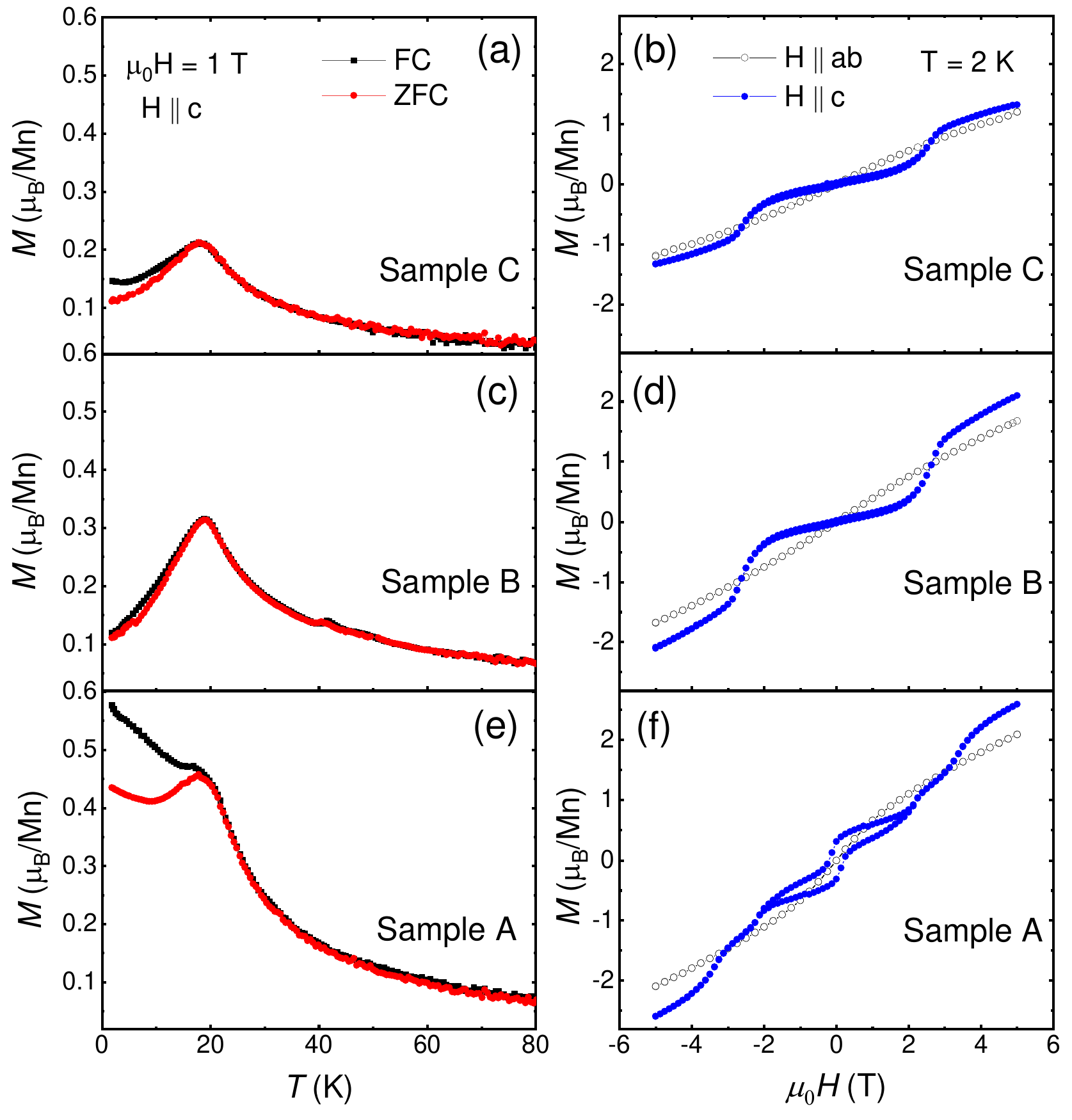}
\caption{Magnetic properties of samples A, B, and C measured by SQUID. Left three panels (a), (c), (e) are temperature dependence of magnetization of the corresponding sample C, B and A, which are measured from 2\,K to 300\,K in the out-of-plane magnetic field of 1\,T after zero-field-cooled (ZFC) or field-cooled (FC) condition. Right three panels (b), (d), (f) are corresponding magnetic field dependence of out-of-plane (solid blue circles) and in-plane (open black circles) magnetization taken at T\,=\,2\,K. The magnetization (M) of the three samples is calculated based on their compositions and thicknesses after subtracting the diamagnetic background signal of the Si substrate.
}
\end{figure}

To investigate the effect of these structural characteristics on the magnetic properties of the layers, we perform a series of temperature and magnetic field dependent magnetization measurements on samples A, B and C using a SQUID magnetometer. The results are presented in Fig.\,3, where background subtraction of the Si substrate contribution is already applied: the M-T curves are corrected by subtracting a substrate background measured from a bare Si substrate under the same conditions, and the M-H curves are corrected by subtracting a linear diamagnetic background. As demonstrated in Fig.\,3(c) and 3(d), sample B (stoichiometric MnBi$_2$Te$_4$) exhibits the characteristics of antiferromagnetic order. The magnetization on warming first reaches its maximum and then drops, forming a kink, which corresponds to the antiferromagnetic transition with a Néel temperature of about 19\,K. The magnetization at 2\,K increases slowly as the out-of-plane field is increased from zero, and exhibits a sharp increase at a field of about 2.5\,T, indicating a spin-flop transition and an out-of-plane easy axis. The absence of hysteresis in the M-H curves is also consistent with the antiferromagnetic state. This magnetic behavior is comparable to that of bulk MnBi$_2$Te$_4$ crystals \cite{2019Nature-MBT,Hosono-Sci.adv,Bulk-MBT}. For the Mn-rich sample C with a mixture of MnBi$_2$Te$_4$ and MnTe phases, Fig.\,3(a) and 3(b) show that similar antiferromagnetic behavior still dominates. The MnTe clusters presented in that sample apparently do not introduce any significant long-range magnetic order. Likely, the MnTe clusters are superparamagnetic or induce frustrated site-disorder in the film, yielding a divergence between zero-field-cooled (ZFC) and field-cooled (FC) magnetic moment at low temperature, as has been observed in other materials \cite{ZFC-FC-cluster,ZFC-FC-spinclass1,ZFC-FC-spinclass2}. For the Mn-poor sample A, with a mixture of MnBi$_2$Te$_4$ and MnBi$_4$Te$_7$ phases, an additional ferromagnetic contribution is present: we observe an increased moment in the FC magnetization and a hysteresis loop in the out-of-plane field at 2\,K, as shown in Fig.\,3(e) and 3(f). This ferromagnetic state is likely related to a ferromagnetic MnBi$_4$Te$_7$ phase, which has a characteristic hysteresis loop saturating at a low field of about 0.2\,T~\cite{Hosono-Sci.adv,ChenXH-PRB,MnBi4Te7}. The unsaturated magnetization at high fields is probably related to the MnBi$_2$Te$_4$ phase.

We additionally perform magnetotransport measurements to further examine the impact of antiferromagnetic order in our MnBi$_2$Te$_4$ films. A piece of sample B is patterned into a six-terminal Hall bar with a size of 200\,$\mu$m$\times$600\,$\mu$m using standard optical lithography methods. Fig.\,4(a) and 4(b) display the temperature dependence of the longitudinal resistivity and its first derivative at various magnetic fields. The first derivative data clearly show that the antiferromagnetic transition appears as an anomaly in the resistivity, occurring around 19\,K in both the 0 and 1\,T curves, consistent with the M-T measurement shown in Fig.\,3(c). This observation can be explained by spin-fluctuation-driven scattering in MnBi$_2$Te$_4$ \cite{spin-fluctuation}. As the temperature approaches the antiferromagnetic ordering temperature, spin scattering affects the transport properties due to the enhanced spin fluctuations. By increasing the magnetic field above 4\,T to overcome the interlayer antiferromagnetic coupling, spin scattering is suppressed and the anomaly disappears. 

Further transport results are provided in Fig.\,4(c) and 4(d), showing the longitudinal resistivity ($\rho$$_{xx}$) and Hall resistivity ($\rho$$_{xy}$) in a perpendicular-to-plane magnetic field. Kinks in both $\rho$$_{xx}$ and $\rho$$_{xy}$ curves occur at about 2.5\,T and disappear above 20\,K. Both values agree with those of the spin-flop transition field and Néel temperature determined by our SQUID results. Moreover, we find a resistivity $\rho$$_{xx}$\,=\,1.6\,m$\Omega$$\cdot$cm, which is comparable to the values of 0.7 and 1.5\,m$\Omega$$\cdot$cm obtained for bulk MnBi$_2$Te$_4$ crystals from which exfoliated flakes of a few 7Ls thickness have been reported to show quantum transport behavior \cite{2020YYWang,MBT-QAHEScience}. 

\begin{figure}
\includegraphics[width=8.5 cm]{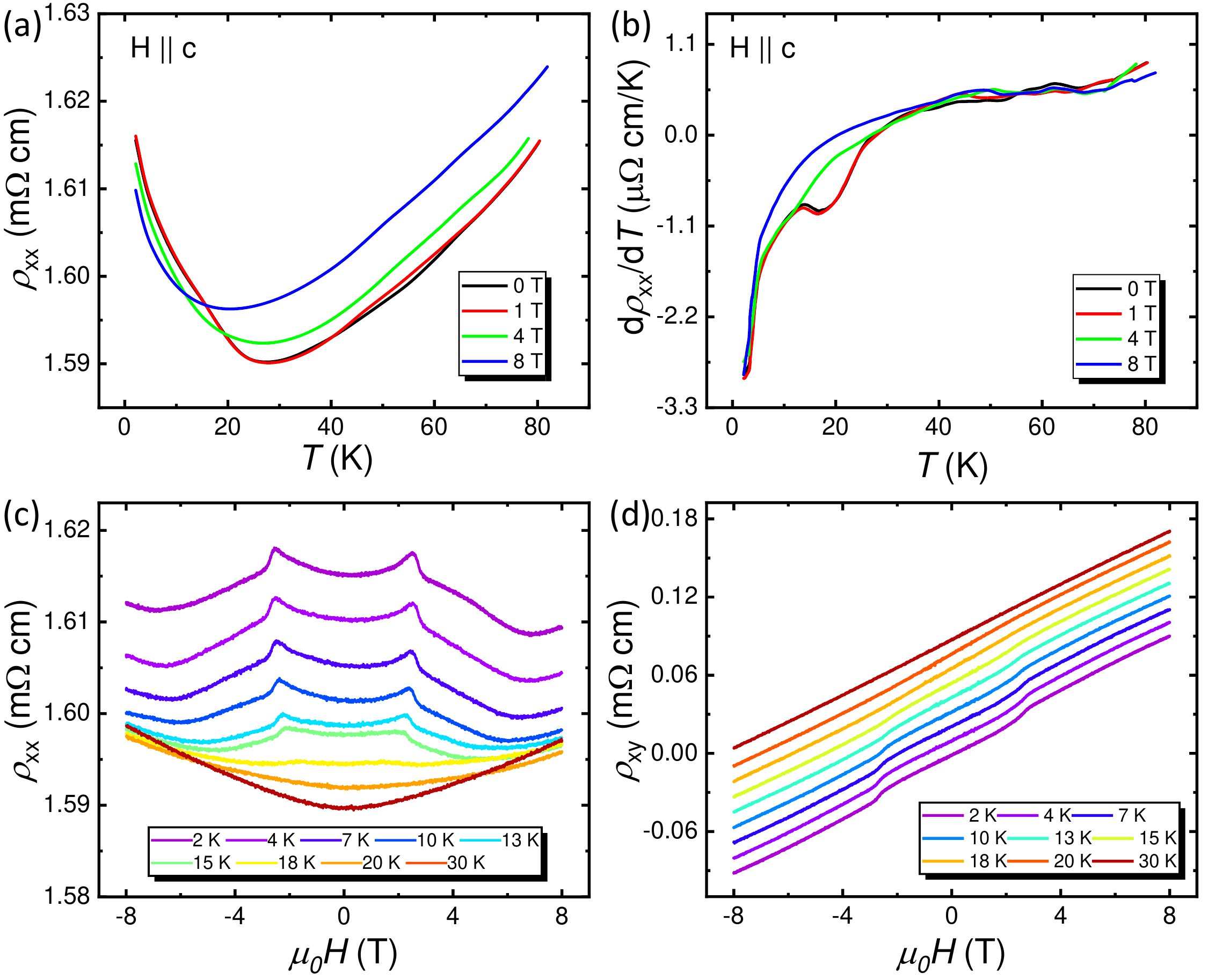}
\caption{Transport characterization of sample B with a thickness of 110\,nm. (a) Longitudinal resistivity ($\rho$$_{xx}$) and (b) its first derivative (d$\rho$$_{xx}$/dT) as a function of temperature at constant magnetic fields of 0, 1, 4, 8\,T. (c) Longitudinal resistivity ($\rho$$_{xx}$) and (d) Hall resistivity ($\rho$$_{xy}$) measurements at various temperatures from 2\,K to 30\,K. $\rho$$_{xy}$ curves are offset for clarity. All data are taken in a magnetic field perpendicular to the plane.
}
\end{figure}

In summary, we have realized high-quality stoichiometric MnBi$_2$Te$_4$ films grown on wafer-sized Si(111) substrates by molecular beam epitaxy. A large layer thickness of about 100\,nm allows us to determine the composition, structural phases, defects, and magnetic properties with high accuracy. The layer properties are well controlled by MBE parameters. The EDS and XRD analysis show high-quality layers with the stoichiometry of MnBi$_2$Te$_4$. TEM reveals large regions with perfect MnBi$_2$Te$_4$ layer structure and extended defects which are typical for vdW epitaxy. The magnetic order is antiferromagnetic as in bulk crystals with slightly reduced Néel temperature and spin-flop transition field. Magnetometry indicates ferromagnetic contributions in Mn-poor layers, while Mn-rich layers with regions of a MnTe phase may be indicated by a difference between ZFC and FC magnetic moment occurring at low temperatures. Because of the large thickness, which is necessary for proper EDS, XRD and SQUID characterization, our MnBi$_2$Te$_4$ films are highly conducting and cannot be gated into the QAHE state. The well-controlled MBE growth of stoichiometric  MnBi$_2$Te$_4$ films presented in this study does however provide the necessary growth know-how for growing both high-quality few 7Ls films, as well as Mn(Bi$_{x}$Sb$_{1-x}$)$_2$Te$_4$ alloys. Based on our studies, both approaches appear promising to achieve the insulating bulk necessary to observe the QAHE {\cite{corbino-paper}.

\begin{acknowledgements} We gratefully acknowledge the financial support of the Free State of Bavaria (the Institute for Topological Insulators), the Deutsche Forschungsgemeinschaft (SFB 1170, 258499086), and the European Commission under the H2020 FETPROACT Grant TOCHA (824140).
\end{acknowledgements}

%\section*{Data Availability}
%The data that support the findings of this study are available from the corresponding author upon reasonable request.

%\begin{appendix}

%\end{appendix}

\end{document}